\documentclass[twocolumn, 10pt, amssymb, aps, physrev, longbibliography]{revtex4-2}

\usepackage{graphicx}
\graphicspath{ {graphics/} }
\usepackage{amsmath}
\usepackage{siunitx}

\usepackage[utf8]{inputenc}

\usepackage[colorlinks=true, linkcolor=black, citecolor=black, urlcolor=black, unicode=true]{hyperref}

\newcommand\I{\mathrm{i}}
\newcommand\di{\mathrm{d}}
\newcommand\hc{\mathrm{h.c.}}
\DeclareMathOperator{\tridiag}{tridiag}

\begin{document}

\title{Interaction-induced directional transport on periodically driven chains}

\author{Helena Drüeke and Dieter Bauer}
\address{Institute of Physics, University of Rostock, 18051 Rostock, Germany}
\email{helena.drueeke@uni-rostock.de, dieter.bauer@uni-rostock.de}

\begin{abstract}
We study a driven system in which interaction between particles causes their directional, coupled movement.
In that model system, two particles move alternatingly in time on two coupled chains.
Without interaction, both particles diffuse along their respective chains, independent from one another.
Interaction between them, no matter if attractive or repellent, leads to an energetic separation of configurations where the particles are close to each other and those where they are farther separated.
The energy difference causes close-by particles to remain bound together, forming a doublon.
Their relative position in the starting configuration determines whether the doublon moves to the left or right or remains stationary due to the periodic driving.
\end{abstract}

\maketitle

\section{Introduction}
\label{sec:introduction}
Directional transport in physical systems can be achieved in various ways.
The most obvious one is applying an external field, e.g., an electric field that accelerates a charged particle in a particular direction. An alternating electric field can also lead to directional transport. A simple example is an electron emitted at, say,  $t = 0$ into a linearly polarized laser field, e.g., by ionization.
Depending on the emission time, the electron may drift in opposite directions, parallel to the polarization of the incident laser field. Other ways to achieve directional transport are by topologically protected edge currents through the breaking of time-reversal symmetry, e.g., by a magnetic field or spin-orbit coupling (Hall effect(s)~\cite{hall_new_1879, dyakonov_current-induced_1971, hofstadter_energy_1976, von_klitzing_40_2020}), or by periodic driving and asymmetric potentials ((semi)classical~\cite{reimann_brownian_2002, hanggi_artificial_2009} and quantum ratchets~\cite{yukawa_quantum_1997, reimann_quantum_1997}).
Interactions between the particles will affect the particle dynamics, but as long as the particle interaction is symmetric under particle exchange, one would not expect directional transport to arise. However, in this work, we present a minimal model of a driven two-particle system that shows directional transport due to interaction, even though this interaction is symmetric under particle exchange.
Moreover, the drive is spatially symmetric (unlike the laser example above), and no asymmetric potentials are involved (in contrast to the ratchet systems).
Instead, the key to directional transport in our system is the alternating driving of the two particles.

While the interaction is always on in our model system, the hopping of each particle is only allowed for half of the driving period.
In this case, the initial configuration determines in which direction the bound pair of particles (i.e., doublon) moves.
The doublon does not exist without interaction, and the two particles simply diffuse without preferred directionality.
The alternating drive where only one of the two particles is allowed to move per half period implies that the two particles are distinguishable and should be independently addressable by external fields.
While such quantum systems probably cannot be found in nature, synthetic models exist, such as ultracold atoms in optical lattices~\cite{holthaus_floquet_2015, nakajima_topological_2016, lohse_thouless_2016, fujiwara_transport_2019} 
or photonic waveguides~\cite{rechtsman_photonic_2013, lumer_self-localized_2013, ablowitz_linear_2014,leykam_edge_2016, mukherjee_observation_2020, mukherjee_observation_2021, jurgensen_quantized_2023}.

The paper consists of the following parts: We introduce the model in Sec.~\ref{sec:system} and explore the behavior of one particle during half its driving period in Sec.~\ref{sec:single}. The doublon dynamics can be conveniently analyzed by mapping onto a 2D system, as discussed in Sec.~\ref{sec:2D}. Finally, we conclude and give an outlook in Sec.\ \ref{sec:concl}.

Throughout the paper, we use units in which $\hbar=1$.

\section{System}
\label{sec:system}
\begin{figure}[hbt]
\centering
\includegraphics{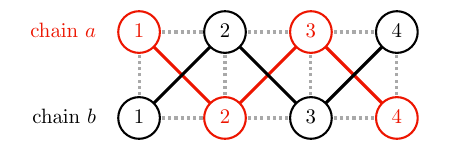}
\caption{Chains $a$ and $b$ of identical length $N = 4$ (we chose this small $N$ for illustration purposes, but performed all calculations with much longer chains).
The red and black lines indicate the hopping $J$ of particles $a$ and $b$ on their respective chains.
Dashed gray lines indicate the interaction $V$ between nearest-neighbor sites on different chains.
}
\label{fig:interlayerstrip}
\end{figure}
We consider the lattice shown in Fig.~\ref{fig:interlayerstrip}, consisting of two chains $a$ and $b$ of length $N$ with one particle on each chain (also labeled  $a$ and $b$).
Each particle may hop along its respective chain; hoppings to the other chain are prohibited.
The particles move alternatingly, starting with particle $a$.
The interaction between particles is between nearest neighbors, i.e., across the chains.

The Hamiltonian reads
\begin{equation}
\begin{aligned}
\hat{H}(t) =& \sum_{\langle i,j\rangle} \left(J_a(t) \hat{a}^\dagger_{i} \hat{a}_j + J_b(t) \hat{b}^\dagger_{i} \hat{b}_j\right) \\
&+  V \sum_{\langle\langle i,j \rangle\rangle} \hat{n}^{(a)}_{i} \hat{n}^{(b)}_{j},
\end{aligned}
\end{equation}
where $\hat{a}$ and $\hat{b}$ are annihilation operators on chains $a$ and $b$, respectively, $\hat{a}^\dagger$ and $\hat{b}^\dagger$ are the corresponding creation operators, and $\hat{n}^{(a)}_i = \hat{a}^\dagger_i \hat{a}_i$ and $\hat{n}^{(b)}_j = \hat{b}^\dagger_j \hat{b}_j$ are the occupation number operators.
$\langle i,j\rangle$ indicates nearest neighbors within a chain, $\langle\langle i,j \rangle\rangle$ nearest neighbors across the chains.

The hoppings  $J_{a,b}(t)$ are assumed to be periodic with a period $T$ and piece-wise constant,
\begin{subequations}
\begin{align}
J_a(t) &=
\begin{cases}
J & 0 \leq t < T/2 \\
0 & T/2 \leq t < T
\end{cases}
\\
J_b(t) &=
\begin{cases}
0 & 0 \leq t < T/2 \\
J & T/2 \leq t < T.
\end{cases}
\end{align}
\end{subequations}
We set $J = 1$ in all plots throughout this publication.
With the labelling in Fig.~\ref{fig:interlayerstrip}, we can write
\begin{equation}
\begin{aligned}
\hat{H}(t) = & \sum_{i=1}^{N-1} \bigg(\left(J_a(t) \hat{a}^\dagger_{i} \hat{a}_{i+1} + J_b(t) \hat{b}^\dagger_{i} \hat{b}_{i+1}\right) + \mathrm{h.c.} \\
& \phantom{\sum_{i=1}^{N-1} \bigg(} + V \left(\hat{n}^{(a)}_{i} \hat{n}^{(b)}_{i+1} + \hat{n}^{(a)}_{i+1} \hat{n}^{(b)}_{i}\right) \bigg)\\
& + V \sum_{i=1}^{N}\hat{n}^{(a)}_{i} \hat{n}^{(b)}_{i}
\end{aligned}
\end{equation}

\section{Movement of one particle during a half period}
\label{sec:single}
We investigate particle $a$'s movement on its chain $a$ during the first half-period ($0 \leq t < T/2$).
Particle $a$ starts in site $i$ and propagates.
Particle $b$ is located in site $j$ and remains stationary during this time.

The Hamiltonian during this phase
\begin{equation}
\hat{H}
= \hat{H}_J + \hat{H}_V
\label{eq:ham_J+V}
\end{equation}
consists of two parts, one describing the hopping
\begin{equation}
\hat{H}_J
=
\tridiag(J, 0, J)
\end{equation}
and one describing the interaction
\begin{equation}
\hat{H}_V
=
\left(v_{k, l}\right)
\end{equation}
\begin{equation}
v_{k, l}
=
\begin{cases}
V & k = l = j - 1, j, j + 1 \\
0 & \mathrm{else}
\end{cases}
\end{equation}
on sites neighboring the position $j$ of particle $b$.

\subsection{\texorpdfstring{$V \gg J$}{V>>J}}
\label{subsec:single_VggJ}
Assuming $|i - j| \le 1$ with a strong potential $V \gg J$ confines particle $a$ to the three sites $j-1$, $j$, and $j+1$ due to the energetic separation of these states from the others.
The $N \times N$ Hamiltonian~\eqref{eq:ham_J+V} becomes limited to these three states ($3 \times 3$),
\begin{equation}
\hat{H}
=
\begin{pmatrix}
V & J & 0 \\
J & V & J \\
0 & J & V
\end{pmatrix}
\end{equation}
with eigenenergies
\begin{equation}
E_0     = V, \quad
E_{1,2} = V \pm \sqrt{2} J
\end{equation}
and eigenstates
\begin{equation}
\varphi_0 =
\begin{pmatrix}
1 \\
0 \\
-1
\end{pmatrix}, \quad
\varphi_{1,2} =
\begin{pmatrix}
1 \\
\pm \sqrt{2} \\
1
\end{pmatrix}.
\end{equation}
We can now write any time-dependent state as
\begin{equation}
\psi(t) = \sum_{k=0}^2 c_k \exp(- \I E_k t) \varphi_k.
\end{equation}

\begin{figure}[hbt]
\centering
\centering
\includegraphics{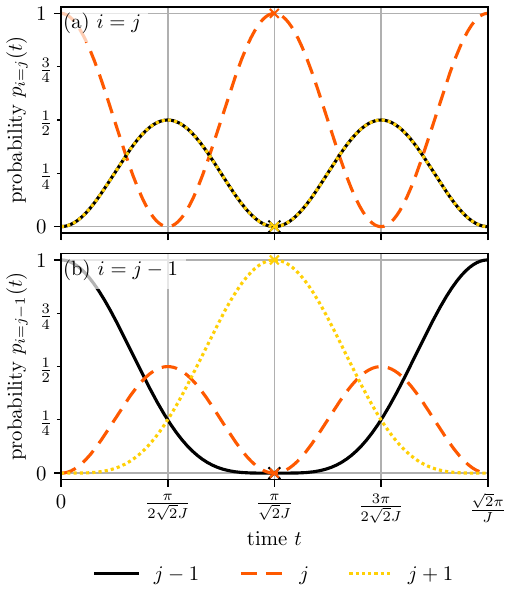}
\caption{
Probabilities of particle $a$ as a function of time with $V \gg J$
(a) for starting position $i = j$ and
(b) for starting position $i = j - 1$.
The crosses mark the probabilities at the end of the driving phase $t_a = \frac{\pi}{\sqrt{2} J}$.
}
\label{fig:three_sites}
\end{figure}

\subsubsection{\texorpdfstring{$i = j$}{i=j}}
If particle $a$ starts at $i = j$, $\psi_{i=j}(0) = (0, 1, 0)^\mathsf{T}$, the coefficients are $c_0 = 0$ and $c_{1,2} = \pm \frac{1}{2 \sqrt{2}}$, resulting in
\begin{equation}
\psi_{i=j}(t) = \frac{\exp(-\I V t)}{\sqrt{2} \I}
\begin{pmatrix}
\sin\left(\sqrt{2} J t\right) \\
\sqrt{2} \I \cos\left(\sqrt{2} J t\right) \\
\sin\left(\sqrt{2} J t\right)
\end{pmatrix}.
\end{equation}
The probability is
\begin{equation}
p_{i=j}(t)
= |\psi_{i=j}(t)|^2
= \frac{1}{2}
\begin{pmatrix}
\sin^2\left(\sqrt{2} J t\right) \\
2 \cos^2\left(\sqrt{2} J t\right) \\
\sin^2\left(\sqrt{2} J t\right)
\end{pmatrix},
\end{equation}
shown in Fig.~\ref{fig:three_sites}(a).
The particle moves symmetrically from the starting site $j$ to the left and right neighbors $j \pm 1$, where it reaches a maximum probability of $0.5$ at time $t = \frac{\pi}{2 \sqrt{2} J}$ before completely returning to site $j$ at $t = \frac{\pi}{\sqrt{2} J}$.

\subsubsection{\texorpdfstring{$i = j - 1$}{i=j-1}}
If particle $a$ starts at $i = j - 1$, $\psi_{i=j-1}(0) = (1, 0, 0)^\mathsf{T}$, the coefficients are $c_0 = \frac{1}{2}$ and $c_{1,2} = \frac{1}{4}$, resulting in
\begin{equation}
\psi_{i=j-1}(t) = \frac{\exp(-\I V t)}{2}
\begin{pmatrix}
1 + \cos\left(\sqrt{2} J t\right) \\
-\sqrt{2} \I \sin\left(\sqrt{2} J t\right) \\
-1 + \cos\left(\sqrt{2} J t\right)
\end{pmatrix}.
\end{equation}
The probability is
\begin{equation}
p_{i=j-1}(t)
=
\begin{pmatrix}
\cos^4\left(J t / \sqrt{2}\right) \\
\sin^2\left(\sqrt{2} J t\right) / 2 \\
\sin^4\left(J t / \sqrt{2}\right)
\end{pmatrix},
\end{equation}
shown in Fig.~\ref{fig:three_sites}(b).

\begin{figure}[hbt]
\centering
\includegraphics{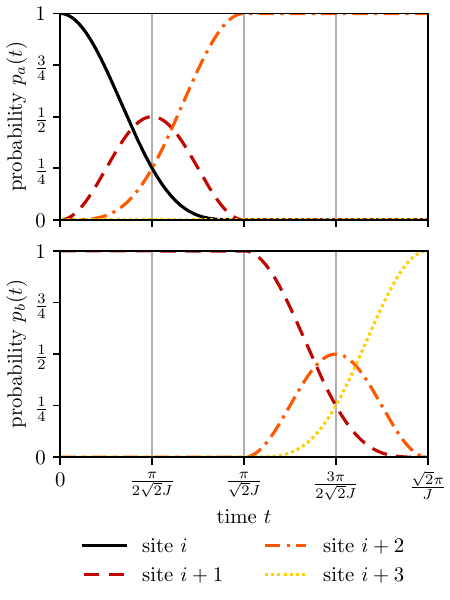}
\caption{
Probabilities of particles $a$ and $b$ as a function of time.
Particle $a$ moves from site $i$ via site $i + 1$ to site $i + 2$ during the first phase, then particle $b$ moves from site $i + 1$ via site $i + 2$ to site $i + 3$ during the second phase.
}
\label{fig:leapfrog}
\end{figure}

\begin{figure}[hbt]
\centering
\includegraphics{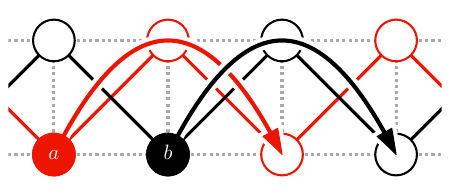}
\caption{
Sketch of the leapfrogging movement of particles $a$ and $b$ during a complete driving cycle.
During the first phase ($0 \leq t < T/2$), particle $a$ jumps over particle $b$ and two sites to the left.
Then, during the second phase ($T/2 \leq t < T$), particle $b$ jumps over particle $a$ and two sites to the left.
}
\label{fig:leapfrog_sketch}
\end{figure}

We choose $t_a = \frac{\pi}{\sqrt{2} J}$ to achieve a complete transfer of particle $a$ from site $j - 1$ to site $j + 1$.
Particle $a$ leapfrogs over particle $b$ from its left to right neighbor.
If we choose the timing of the second phase of the driving cycle as $t_b = \frac{\pi}{\sqrt{2} J}$, particle $b$ will leapfrog over particle $a$, leading to directional transport.
Effectively, both particles move two sites to the right without spreading.
Fig.~\ref{fig:leapfrog} shows the probabilities $p_a(t)$ and $p_b(t)$ for the complete cycle.
Fig.~\ref{fig:leapfrog_sketch} shows a sketch of the particles' movement.

\subsubsection{\texorpdfstring{$i = j + 1$}{i=j+1}}
If particle $a$ starts at $i = j + 1$, $\psi_{i=j+1}(0) = (0, 0, 1)^\mathsf{T}$, it will analogously leapfrog over particle $b$ to site $j - 1$, resulting in directional transport to the left.

\subsection{\texorpdfstring{$V = 0$}{V=0}}
\begin{figure}[hbt]
\centering
\includegraphics{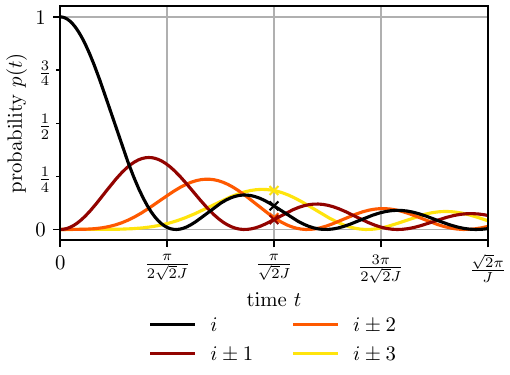}
\caption{
Probabilities of particle $a$ in different sites as a function of time with $V = 0$.
The crosses mark the probabilities at the end of the driving phase $t_a = \frac{\pi}{\sqrt{2} J}$.
}
\label{fig:spreading}
\end{figure}

For potential $V = 0$, the position $j$ of particle $b$ does not influence particle $a$'s movement.
The Hamiltonian~\eqref{eq:ham_J+V} simplifies to
\begin{equation}
\hat{H}
= \hat{H}_J
= \tridiag(J, 0, J)
\end{equation}
As shown in Fig.~\ref{fig:spreading}, particle $a$ spreads symmetrically to the left and the right.

\subsection{\texorpdfstring{$V \ne 0$}{V!=0}}
For potential $V \ne 0$ but not $V \gg J$, we must use the whole Hamiltonian~\eqref{eq:ham_J+V} to describe the system.

\begin{figure}[hbt]
\centering
\includegraphics{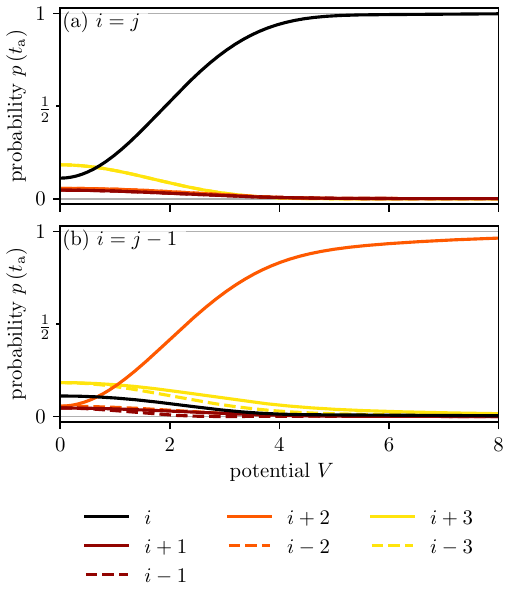}
\caption{
Probabilities of particle $a$ at time $t_a = \frac{\pi}{\sqrt{2} J}$ as a function of interaction $V$
(a) for starting position $i = j$ and
(b) for starting position $i = j - 1$.
}
\label{fig:after_phase}
\end{figure}

A video in the supplemental material shows the evolution of the probabilities for increasing interaction $V$ going from the spreading at $V = 0$ shown in Fig.~\ref{fig:spreading} to the periodic returns at $V \gg J$ shown in Fig.~\ref{fig:three_sites}.
We are mainly interested in the probabilities at the end of phase $a$, $t_a = \frac{\pi}{\sqrt{2} J}$.
These are marked by crosses in Figs.~\ref{fig:three_sites}~and~\ref{fig:spreading}.
Fig.~\ref{fig:after_phase} shows the probabilities $p(t_a)$ as a function of the interaction $V$.
Even at relatively small interactions $V \gtrapprox 6$, the initial configuration $i = j$ remains stationary, $p_{i = j}(t_a) \approx 1$.
The leapfrogging state (starting at $i = j \pm 1$) needs higher interaction strengths $V \gtrapprox 20$ to remain localized ($p_{j \mp 1}(t_a) \approx 1$) while jumping from site $j \pm 1$ to site $j \mp 1$.

\section{Mapping to 2D}
\label{sec:2D}
\begin{figure}[hbt]
\centering
\includegraphics{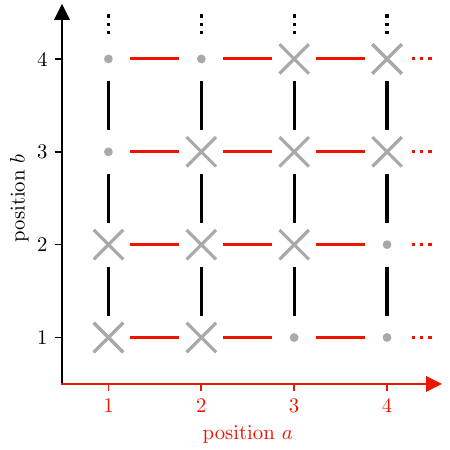}
\caption{
Mapping of the Hamiltonian to a 2D lattice with the two particles' indices along the two axes ($a$-chain index $s$ at the $x$ axis, $b$-chain index $t$ at the $y$ axis).
Red and black lines connecting sites indicate the hoppings $J_a(t)$ and $J_b(t)$.
Gray crosses indicate the combinations of lattice sites for which the interaction potential is non-vanishing, i.e., $(s,t)=(1,1),(1,2),(2,1),(2,2),(2,3),(3,2),(3,3), \ldots$ .
}
\label{fig:map2D}
\end{figure}
We map the two chains to a square grid, as shown in Fig.~\ref{fig:map2D}.
The positions of particles $a$ and $b$  are plotted along the horizontal and vertical directions, respectively.

\subsection{Interacting subsystem for \texorpdfstring{$V \gg J$}{V>>J}}
\begin{figure}[hbt]
\centering
\includegraphics{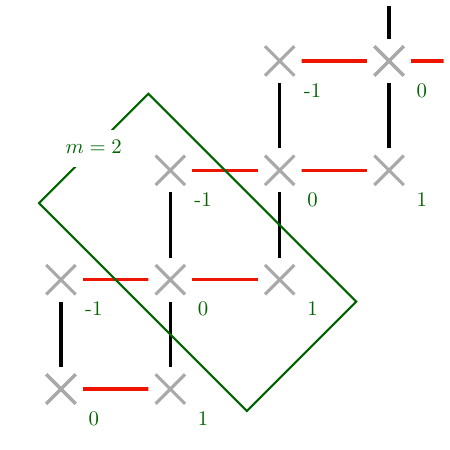}
\caption{
Lattice on which the doublon dynamics takes place if $V\gg J_{a,b}$, with new labeling and the unit cell indicated in green.
}
\label{fig:doublons_unit}
\end{figure}

For strong interactions $V \gg J$, interacting states (located on sites marked by crosses in Fig.~\ref{fig:map2D}) are energetically separated from non-interacting states (located on sites marked by dots).
If the initial state is interacting, it will remain an interacting state.
Hence, for $V \gg J$, we only need to consider a subset of the 2D system, as shown in Fig.~\ref{fig:doublons_unit}.
The unit cell $m$ contains three sites, labeled by the difference of positions $a$ and $b$: $1$, $0$, and $-1$.
This reduced system is quasi-1D, effectively a three-site wide ribbon.

\subsubsection{Stationary states}
A state initially located at site $(m, 0)$ will split towards sites $(m - 1, -1)$ and $(m, 1)$ during the first phase, returning to $(m, 0)$ at the end of the phase, $t_a = \frac{\pi}{\sqrt{2} J}$. 
During the second phase, it will equivalently split towards sites $(m, -1)$ and $(m - 1, 1)$ before returning to $(m, 0)$ at the end of the driving cycle $T = t_a + t_b = \frac{\sqrt{2} \pi}{J}$.
The state appears to be stationary when looking 
stroboscopically after complete driving cycles.

\subsubsection{Leapfrogging states}
A state starting in site $(m, \pm 1)$ moves to site $(m \mp 1, \mp 1)$ during the first phase and then to site $(m \mp 2, \pm 1)$ during the second phase.
The states move two unit cells in each cycle.

\subsubsection{Reflection at the corner}
The two preceding paragraphs described the evolution of states in an infinite system or the bulk of finite chains.
Now, we will investigate the effects of borders.
Fig.~\ref{fig:doublons_unit} shows the bottom left corner, with the complete unit cell $m = 1$.
The upper right corner is a partial unit cell $m = N$, containing only the site $(N, 0)$ with
sites $(N, \pm 1)$ absent.

For the Hamiltonian at the edge, one needs to consider only two sites during each driving phase (instead of three for the bulk),
\begin{equation}
\hat{H}
=
\begin{pmatrix}
V & J \\
J & V
\end{pmatrix}.
\end{equation}
The eigenenergies are
\begin{equation}
E_{1,2} = V \pm J,
\end{equation}
and the eigenstates are
\begin{equation}
\varphi_{1,2} =
\begin{pmatrix}
1 \\
\pm 1
\end{pmatrix}.
\end{equation}
We can now write any time-dependent state during that driving phase as
\begin{equation}
\psi(t) = \sum_{k=1}^2 c_k \exp(- \I E_k t) \varphi_k.
\end{equation}
Without loss of generality, we initialize the state as $\psi(0) = (1, 0)^\mathsf{T}$.
The coefficients become $c_1 = c_2 = 1 / 2$, resulting in
\begin{equation}
\psi(t) = \exp(- \I V t)
\begin{pmatrix}
\cos(J t) \\
- \I \sin(J t)
\end{pmatrix}
\end{equation}
and the probability
\begin{equation}
p(t)
= |\psi(t)|^2
=
\begin{pmatrix}
\cos^2(J t) \\
\sin^2(J t)
\end{pmatrix}.
\end{equation}
Compared to the three-site Hamiltonian in section~\ref{sec:single}, the oscillation frequency of the two-site Hamiltonian is decreased from $\sqrt{2} J$ to $J$.
Therefore, at the end of the phase $t_a = \frac{\pi}{\sqrt{2} J}$, the state is incompletely transferred from one site to the next.
\begin{equation}
p\left(t_a\right)
=
\begin{pmatrix}
\cos^2\left(\pi / \sqrt{2}\right) \\
\sin^2\left(\pi / \sqrt{2}\right)
\end{pmatrix}
\approx
\begin{pmatrix}
0.3669 \\
0.6331
\end{pmatrix}.
\end{equation}
The corner influences the stationary state starting at site $(1, 0)$.
It leaks into $(1, 1)$ in the first phase, from where it continues to $(2, -1)$ in the second phase.
It also leaks into $(1, -1)$ in the second phase.
The stationary state sends out leapfrogging states until it vanishes.
Here, we described the edge at $m = 1$, but the behavior at the other edge is equivalent.

The leapfrogging states split up when they run into an edge, similar to the stationary states.

\subsubsection{Interpretation as a spin-1 system}
Labeling sites in the unit cell as $-1$, $0$, and $1$ already suggests an analogy to a spin-1 system.
The leapfrogging states undergo a spin-flip operation from $\pm 1$ to $\mp 1$ in each phase, accompanied by a spatial movement.
The spin $0$ states are unaffected by the spin-flip and remain in the same location.
Although there is a similarity to the quantum spin Hall effect in the sense that the transport direction depends on spin, there are essential differences. Besides the third spin-degree of freedom $0$ without transport, the spin flips during transport in our model system.

\subsection{Band structure}
\begin{figure}[hbt]
\centering
\includegraphics{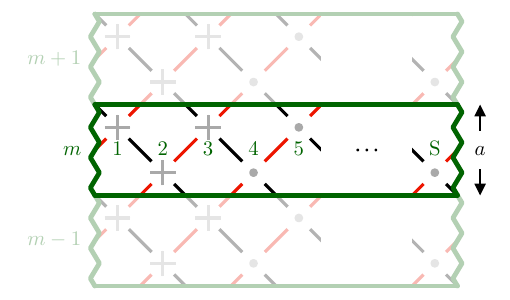}
\caption{
Unit cell of the $\ang{45}$-rotated system.
The sites within it are numbered by $\alpha=1,2,3,\ldots,S$ with even $S$.
The crosses mark the diagonal sites, and the dots mark the non-diagonal sites.
The unit cell is infinitely repeated in the vertical direction and numbered by the index $m$.
The height of the unit cell, $a$, is marked.
Periodic boundaries are employed horizontally, connecting sites $S$ and $1$ within the same unit cell.
}
\label{fig:sketch_dia}
\end{figure}

To calculate a band structure, we use a unit cell (shown in Fig.~\ref{fig:sketch_dia}) which contains non-diagonal sites in addition to the three diagonal sites.
The sites are numbered $\alpha=1,2,3,\ldots,S$ with even $S$.
The unit cell is repeated infinitely in one direction and numbered by an index $m$.
We employ periodic boundary conditions in the other, finite direction, connecting the left and right edges of the unit cell.
While this periodicity does not exist in the complete 2D system, the alternative would create diagonal edges, which do not exist in the 2D square system since there are only horizontal and vertical edges.
The edge states at these diagonal edges would obfuscate the bands we are interested in.

We can write the Hamiltonians for the two phases of the driving cycle in real space as
\begin{equation}
\begin{aligned}
\hat{H}_i = \sum_m \Bigg(& J \sum_{\alpha\ \textrm{odd}} \left(\hat{h}_i(m, \alpha) + \hc\right) \\
                         &+ V \sum_{\alpha=1}^3 | m, \alpha \rangle \langle m, \alpha |\Bigg)
\end{aligned}
\end{equation}
with
\begin{equation}
\begin{aligned}
\hat{h}_a(m, \alpha) =\ &   | m, \alpha \rangle \langle m, (\alpha-1) \mod S | \\
                        & + | m, \alpha \rangle \langle m+1, (\alpha+1) \mod S | \\
\hat{h}_b(m, \alpha) =\ &   | m, \alpha \rangle \langle m, (\alpha+1) \mod S | \\
                        & + | m, \alpha \rangle \langle m+1, (\alpha-1) \mod S |.
\end{aligned}
\end{equation}
We transform the Hamiltonians to $k$-space by making the Bloch ansatz~\cite{bloch_uber_1929}
\begin{equation}
| m, \alpha \rangle = \frac{a}{2\pi} \int_\mathrm{BZ} \di k \, \exp(-\I  k m a) \, | k, \alpha \rangle,
\end{equation}
where $a$ is the lattice constant in the vertical direction in Fig. \ref{fig:sketch_dia}.
We obtain
\begin{equation}
\begin{aligned}
\hat{H}_i = \frac{a}{2\pi}\int_{\mathrm{BZ}} \di k \, | k \rangle \langle k | \Bigg(
& J \sum_{\alpha\ \textrm{odd}} \left(\hat{h}_i(k, \alpha) + \hc\right) \\
&+ V \sum_{\alpha=1}^3 |\alpha \rangle \langle \alpha | 
\Bigg)
\end{aligned}
\end{equation}
with
\begin{equation}
\begin{aligned}
\hat{h}_a(k) &= |\alpha \rangle \langle \alpha-1 | + \exp(\I k a) | \alpha \rangle \langle \alpha+1 | \\
\hat{h}_b(k) &=  |\alpha \rangle \langle \alpha+1 | + \exp(\I k a) | \alpha \rangle \langle \alpha-1 |.
\end{aligned}
\end{equation}
The time evolution operator is (in units where $\hbar=1$)
\begin{equation}
\hat{U}(t)
= \exp\left(\frac{T}{2\I} \hat{H}_b\right) \exp\left(\frac{T}{2\I} \hat{H}_a\right)  .
\end{equation}
Solving the equation
\begin{equation}
\hat{U}(T) \psi_\mathrm{F} = \lambda_\mathrm{F} \psi_\mathrm{F}
\end{equation}
gives the Floquet~\cite{floquet_sur_1883} eigenstates $\psi_\mathrm{F}$, and
the Floquet energies $\varepsilon_\mathrm{F}$ are calculated from the eigenvalues $\lambda_\mathrm{F} = \exp(- \I \varepsilon_\mathrm{F} T)$.

\begin{figure}[hbt]
\centering
\includegraphics{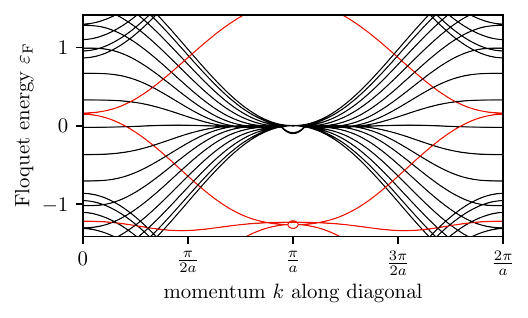}
\caption{
Band structure of a 20-site wide strip with $V = 10$.
The red bands reside on the three sites with the modified potential, and the black bands on other sites.
}
\label{fig:bands}
\end{figure}

The resulting band structure in Fig.~\ref{fig:bands} confirms our previous observations on the behavior of the doublons.
They are located on the three sites $\alpha=1,2,3$, and Floquet eigenstates where this is the case are drawn red in Fig.~\ref{fig:bands}.
One of these doublon bands is quite flat, corresponding to the stationary doublons.
The two sloped bands correspond to doublons moving in opposite directions along the diagonal.
The other bands are shown in black and form a continuum for $N \to \infty$.
These bands are the diffusing states.

Depending on the potential $V$, some diffusing bands have non-zero energy at the center of the Brillouin zone, $\varepsilon_\mathrm{F}\left(k = \frac{\pi}{a}\right) \ne 0$.
These are edge states, localized at the boundary between $\alpha=3$ and $4$, and  between $\alpha=S$ and $1$.

\begin{figure}[hbt]
\centering
\includegraphics{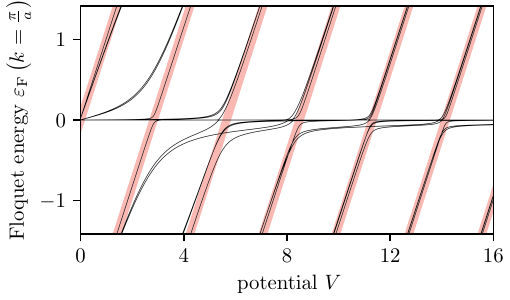}
\caption{
Floquet energies $\varepsilon_\mathrm{F}$ at $k=\frac{\pi}{a}$ as a function of potential $V$ for a 20-site wide strip.
The energies of the doublons are marked by a reddish shadow.}
\label{fig:energies}
\end{figure}

Fig.~\ref{fig:energies} shows the Floquet energies $\varepsilon_\mathrm{F}\left(k=\frac{\pi}{a}\right)$ as a function of potential $V$.
The bulk states are at constant $\varepsilon_\mathrm{F}\left(k=\frac{\pi}{a}\right) = 0$.
The energies of the doublons increase linearly with $V$, as indicated by the reddish shadow $\varepsilon_\mathrm{F} = V$.
The energies of the edge states show an interesting behavior:
They have a tilted pole at $V \approx 3$, where they approach the doublon energies.
At higher potentials, they approach the energy of the bulk states, $\lim_{V \to \infty} \varepsilon_\mathrm{F}\left(k=\frac{\pi}{a}\right) = 0$.
There are crossings between the doublon and edge state energies. We have checked that they are avoided crossings by following the Floquet eigenstates.

\section{Conclusion} \label{sec:concl}
We investigated two particles on two linear chains in a periodic driving scheme and showed how their interaction influences their temporal evolution.
Without interaction, both particles diffuse.
With sufficiently strong interaction, they can form a stationary bound state which remains localized without diffusing.
They can also form non-stationary non-diffusing states, which propagate in a leapfrogging manner.
The relative position of the two particles in the starting configuration determines their behavior.
Observing the evolution of the two particles could allow us to measure the strength of the interaction between them and their initial locations.

A possible extension of the system would be going from linear chains to two-dimensional grids on which the particles move.
The added dimension would enable vertical and diagonal movement of the particles in addition to the horizontal one on the chains.

\bibliography{references.bib}

\end{document}